\documentclass[11pt,preprint,floatfix,nofootinbib,usenatbib]{revtex4-1}
\usepackage[T1]{fontenc}
\usepackage{graphicx}
\usepackage[latin1]{inputenc}
\usepackage{times}
\usepackage{docs}
\usepackage{multirow}
\usepackage{amsmath}
\usepackage{cases}
\usepackage{array}
\usepackage[dvipsnames,usenames]{color}
\usepackage[dvipsnames,usenames]{colortbl}
\usepackage{datetime}
\usepackage{subfigure}
\usepackage{caption}

\newcommand\blfootnote[1]{
  \begingroup
  \renewcommand\thefootnote{}\footnote{#1}
  \addtocounter{footnote}{-1}
  \endgroup
}

\begin{document}
\begin{center}
\section*{\large{EXPLORING A DYNAMICAL PATH FOR $\mathrm{\mathbf{C_2H^-}}$ AND $\mathrm{\mathbf{NCO^-}}$ FORMATION IN DARK MOLECULAR CLOUDS$^\S$}}
\end{center}

\begin{center}
\begin{author}
\author{I. Iskandarov$^1$, F. Carelli$^1$,E. Yurtsever$^2$, R. Wester$^1$ and F.A. Gianturco$^{\ast 1,3}$}

 \blfootnote{ $^*$Corresponding author. Email: Francesco.Gianturco@uibk.ac.at \vspace{6pt}}

\vspace{6pt}
$^1${\em{Institut f\"{u}r Ionen Physik und Angewandte Physik,
Leopold-Franzens-Universit\"{a}t, Technikerstra{\ss}e
25, 6020, Innsbruck, Austria}};

$^2${\em{Department of Chemistry, Koc University, Rumelifeneriyolu, Sariyer 34450, Istanbul, Turkey}};

$^3${\em{Scuola Normale Superiore. Piazza de' Cavalieri 7, Pisa, 56126 Italy}}
\end{author}
\end{center}

%\begin{center}
%\begin{affil}
%\small {Department of Physics, the University of Innsbruck, Technikerstrs. 25/3, A-6020 Innsbruck, Austria}
%\end{affil}
%\end{center}

\section*{Abstract}

This paper deals with the  possible formation of two molecular anions often considered likely components in the physical environments of the Interstellar Medium ( ISM) : $\mathrm{C_2H^-}$ and $\mathrm{NCO^-}$. They are both  discussed here by computationally following the radiative association (RA) mechanism starting from $\mathrm{C_2^-}$, $\mathrm{H}$, $\mathrm{NC^-}$ and $\mathrm{O}$ as partners. The corresponding RA total cross sections produced by the calculations are in turn employed to generate the overall association rates over the relevant range of temperatures. The latter are found to be in line with other molecular ions formed by RA but not large enough to uniquivocally suggest this path as the main route to  the anions formation in the ISM. Other possible paths of formation are also analysed and discussed.
The presence of resonant structures during the association dynamics for both systems is found by the calculations and their consequences are discussed in some detail in the present study.
%\end{abstract}
%\keywords{astrochemistry - molecular data - molecular processes - radiation mechanisms}

\clearpage

%\begin{center}
\section{INTRODUCTION}
%\end{center}

Interstellar space is filled with dilute mixtures of charged particles, atoms and molecules that form dense molecular clouds (DMC)  which  define an important component of the ISM [1] since these molecular environments  are important birthplaces for new stars and for likely new planetary systems. Among the large number of species detected there, the molecular elements containing carbon atom have a very important role in the general chemistry of the ISM and, more specifically for the present study, within the suggested mechanisms for  molecular anions formation.

More than three decades ago, Herbst proposed  the existence of  negatively charged molecules in the interstellar medium, a suggestion which was based on the possible mechanism of dynamical  attachment of low energy electrons to  neutral radicals, followed by radiative stabilization : the Radiative Electron Attachment (REA) process [2]. Later experiments have indeed confirmed the presence of some molecular anions in the ISM  by observing several negatively charged carbon chains of two types: $\mathrm{C_nH^-}$ (n=4,6,8) [3-9] and $\mathrm{C_nN^-}$ (n=1,3,5) [4,7,10-13]. Among these molecular anions the small carbon chain of type $\mathrm{C_2H^-}$ has been only tentatively observed and thus far  its detection has not yet been fully confirmed, as we shall further discuss below [14]. One of the reasons for the $\mathrm{C_2H^-}$ detection difficulty  has been mainly  linked to its disappearance via the competing electron autodetachment channel that makes the REA process not very likely to lead to the final anion's stabilization [15,16], as well as to the experimental findings in laboratory studies of its reaction with $\mathrm{H}$ atoms [17] that suggested the dominance of the electron's  loss path to products, as we shall further discuss below. It was in fact suggested in that work that a physical feature  which is expected  to significantly reduce  the possible formation of $\mathrm{C_2H^-}$ in the ISM by low-energy electron attachment is linked to the low density of the existing vibrational states in the partner molecular radical that would reduce the efficiency of both direct and indirect REA processes[16,18]. One of the goals of the present study is therefore that of  exploring another possible mechanism for the  formation of two of the small Carbon-bearing anions: $\mathrm{C_2H^-}$ and $\mathrm{NCO^-}$ that have been suggested for a long time to be possible components of DMC environments. With the new astronomical observatory ALMA acting in the microwave range, astronomers are expecting more significant findings regarding such special molecular species and therefore the possible clarification of the observational doubts on their actual existence in the ISM [14].

In the present work we investigate the formation of the $\mathrm{C_2H^-}$ and $\mathrm{NCO^-}$ via  dynamical association paths involving spontaneous and stimulated photon emission, given below  in an example for the former species:

\begin{equation}
C_2^-(X ^2\Sigma_g^+) + H(^2S) \rightarrow C_2H^-(^1\Sigma^+) + \hbar\omega
\end{equation}

\begin{equation}
C_2^-(X ^2\Sigma_g^+) + H(^2S) + \hbar\omega \rightarrow C_2H^-(^1\Sigma^+) + 2\hbar\omega
\end{equation}

Thus, in the pseudo- 2D model  dynamical mechanism reported above, as an example  for one of the present molecules, the final anionic systems are stabilized by emitting a photon, either spontaneously or by stimulation through an external stellar photon bath.
  As mentioned above, we have also investigated a molecular anion which could also exist in the ISM: the molecular cyanate $\mathrm{NCO^-}$, because of the rising interest on C-bearing ionic isomers (the latter being in fact the more stable with respect to $\mathrm{ONC^-}$). In the interstellar medium this anion has been known in fact for a long time in condensed phases and  is considered one of the most stable ones in the interstellar icy grain mantles [12]. From the theoretical side, the stability of the present anion has been investigated by computationally  exploring  the lower electronic states involving both  neutral isomers [18]. In the gas phase, the anion has been detected in various halide matrices [13] although, despite its electronic stability  and its large Electron Affinity (EA) value of 3.609 eV [18], the likelyhood of its formation in the ISM remains  an open question. Therefore, it is indeed  interesting to computationally estimate the efficiency of its possible formation path  that would follow the same RA mechanism discussed above. In the present study  we will thus  consider as a mechanism of formation of the $\mathrm{NCO^-}$ anion  the same pseudo-2D  association process that has been mentioned above:

\begin{equation}
NC^-(X ^2\Sigma_g^+) + O(^3P) \rightarrow NCO^-(^1\Sigma^+) + \hbar\omega
\end{equation}

\begin{equation}
NC^-(X ^2\Sigma_g^+) + O(^3P) + \hbar\omega \rightarrow NCO^-(^1\Sigma^+) + 2\hbar\omega
\end{equation}

Therefore the principal aim here shall be that of  evaluating with accurate quantum methods the efficiency of all the above processes in forming the title systems.

The paper is organized as follows: in Section 2, we will briefly outline the computational details involving the quantum structure of the relevant potential energy curves (PECs) for the systems we are studying. In Section 3 we shall initially  provide a brief summary of the formulae employed for the calculations of the relevant cross sections, while afterwards we shall discussin the same ection  their physical features and analyse the role and importance of the resonant processes. The last part of the same section will also discuss the corresponding RA rates under different temperature conditions of the Molecular Clouds. Section 4 shall finally report our discussion of the results obtained and their implication for the formation of both anions in the ISM environments.

\section{COMPUTING THE  $\mathrm{\mathbf{C_2H^-}}$ AND $\mathrm{\mathbf{NCO^-}}$ INTERACTION POTENTIAL CURVES}

In order to obtain the RA cross sections for the present molecular anions, the potential energy curves of the ground electronic states $\mathrm{X ^1\Sigma^+}$ of both $\mathrm{C_2H^-}$ and $\mathrm{NCO^-}$, together with the values and radial behaviour of their electric dipole moments, need to be evaluated. The dipole moments have been computed by  considering the center of mass of the molecules and they are obtained using the same level of theory chosen for the potential energy curves (PECs). The vibrational bound states of the pseudo-2D molecules have also been generated using the LEVEL 8.0 suite of programs [20]. We should further point out here that the atom-molecule RA process we are modelling  is actually a 3D process where all three Jacobi coordinates are in principle involved. However, the diatomic partners possess, chemically speaking, stronger bonds and more rigid structures that those induced by attaching the extra atom during the radiative recombination.Thus, as a first physical approximation to the full process we are initially considering the multiple bonds in anionic partners as being fixed at their equilibrium values and also shall demonstrate from the structure calculations below that the collinear approaches are in both cases the most likely to occur. Hence, the reduction of the initial 3D problem to a pseudo-2D problem was adopted in the present study.

The ground and several excited states of the $\mathrm{C_2H^-}$ had been previously computed  at an accurate level [15]. In their work these authors discussed  the possible approach of atomic  $\mathrm{H}$ towards $\mathrm{C_2^-}$ to form $\mathrm{C_2H^-}$, a process they showed to be occurring only along the lowest electronic state: we one we are considering in the present study. In practice, we further decided to employ a more accurate level of calculation while focussing on the associative processes which follow the lowest electronic states of the partners, in line with what is espected to happen in the dark interstellar clouds of the ISM. It is also interesting to note here that the previous work of ref.[15] has pointed out how the possible recombination paths along the excited PECs of the partners  appear to favour the competing channels of electron autodetachment with formation of the neutral radicals though the Associative Detachment (AD) paths. We shall further return on this aspect of the problem in our final section.

  We have thus performed  computational studies for the  only possible path in which the  $\mathrm{C_2H^-}$ could be formed on its ground state $\mathrm{X ^1\Sigma^+}$ in collision with neutral hydrogen atoms,as it corresponds to the most probable channel for an associative process forming the anionic species. All new calculations have been performed using the MOLPRO suite of computational programmes [21], with which geometries and energies corresponding to the ground electronic state of both $\mathrm{C_2^-}$ and $\mathrm{C_2H^-}$ have been calculated using a Multireference Configuration Interaction (MRCI) with aug-cc-pVTZ basis set. Before the MRCI process, the complete active space CASSCF was defined considering all valence orbitals, whereas the two core orbitals were kept doubly occupied. Furthermore during the process of $\mathrm{H}$ attacking the $\mathrm{C_2^-}$, we kept the $\mathrm{C_2^-}$ bond fixed at its equiliblrium distance $\mathrm{R_{C_2^-}}$ = 1.28 $\mathrm{{\buildrel _{\circ} \over {\mathrm{A}}}}$.This is in line with the expectation that the anionic partner $\mathrm{C_2^-}$ would be existing in a cold environment and therefore in its lowest internal states.  Additionally we will be interested in the collinear process where the $\mathrm{H}$ is approaching along the $\mathrm{C_2^-}$ bond, a dynamical choice that whe have already discussed above and that  will also be analysed in more detail below.

For $\mathrm{NCO^-}$, the radiative association cross sections will be  obtained following the same physical outlook  and computational approach as it was done for $\mathrm{C_2H^-}$. Hence, the ground electronic state $\mathrm{X ^1\Sigma^+}$ calculations have been also performed using MOLPRO package [21], employing second order Moller Plesset perturbation theory (MP2). The electrons of the three atoms $\mathrm{(N, C, O)^-}$ have been represented by the aug-cc-pVQZ basis set. The ground state of $\mathrm{NCO^-}$ is a closed shell structure with  16 valence and 6 core electrons. While calculating the PEC for its lowest electronic configuration, in line with what we have done for the other anionic molecule,  we kept $\mathrm{NC^-}$ bond fixed at its equilibrium geometry $\mathrm{R_{NC^-}}$ = 1.19 $\mathrm{{\buildrel _{\circ} \over {\mathrm{A}}}}$ and varied  the $\mathrm{C-O}$ distance. We also wish to point out that for both the present systems the collinear process turns out to be the most favourable path for the RA channel, as we shall further discuss below in greater detail. Hence our present choice of treating the associative processes as pseudo-2Body processes in which only one additional bond is involved in the present dynamics along the collinear arrangements.

\subsection{THE GROUND ELECTRONIC STATES OF $\mathrm{\mathbf{C_2H^-}}$ AND $\mathrm{\mathbf{NCO^-}}$}
In Figure 1 we report a specific  comparison between our computed PECs for the ground electronic states, and the previously computed PECs obtained from ref. [15], for the case of the $\mathrm{C_2H^-}$ system. Besides reporting  the potential  curves in the main panel, the figure contains of  three additional insets. These smaller panels are presenting an enlarged view of well and of the asymptotic regions of the PECs, where we have compared the analytic extrapolation form of the interaction, dominated by the hydrogen atom dipole polarisability, with the ab initio  points calculated in the long range regions (bottom right):  it can be seen there that the two curves are in very good agreement with each other, thus confirming the good quality of the present calculations. When looking at the region of the potential minimum, we can further see that our computed PEC is markedly deeper by about 2000 $\mathrm{cm^{-1}}$ than the one reported by ref.[15]. The presence of a deeper well will increase the number of bound states supported by the interaction between partners and therefore, as we shall discuss further below, also increase the probability of formation of the complex because more vibrational bound states will be available for the RA stabilization path. The long-range part of the interaction obtained by us is compared with that from earlier calculations in the inset in the upper right part of the figure: the present results are seen to provide a stronger tail to the overall PEC, also causing an increase in the number of the bound states closer to dissociation.
To provide more numerical details, we further report in Table 1  the vibrational levels for the $\mathrm{X ^1\Sigma^+}$ state of $\mathrm{C_2H^-}$, and the values of the Zero Point Energy (ZPE) collocation, coming from both sets of calculations.
\begin{figure}[h]
\includegraphics[width=0.9\textwidth, height=260px]{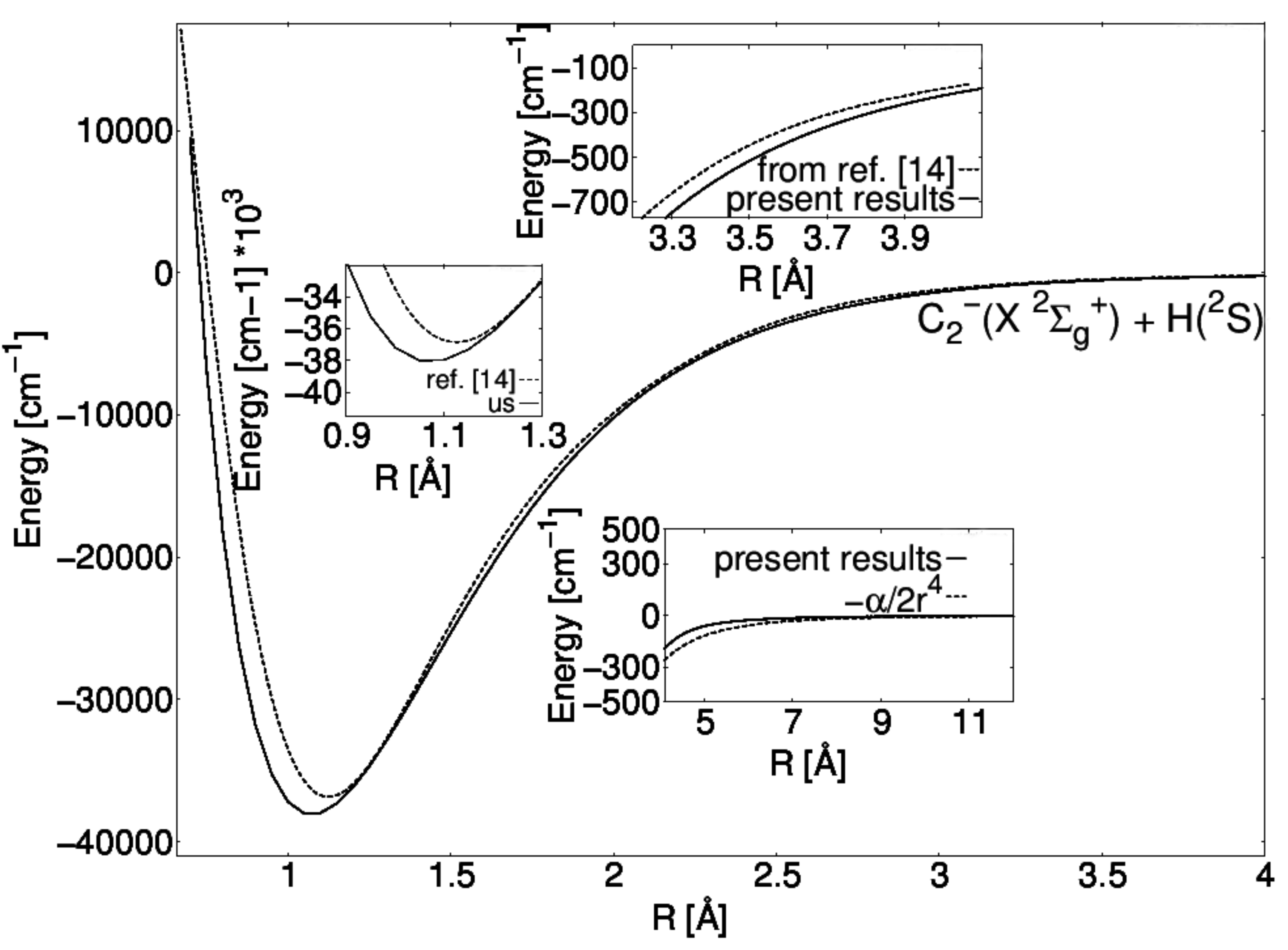}
\caption{Comparison between our present calculations and the earlier  PEC from reference [15] for the ground electronic state of $\mathrm{C_2H^-}$. Here $\mathrm{\alpha}$ = 0.67 $\mathrm{{\buildrel _{\circ} \over {\mathrm{A}}}^3}$ is the dipolar polarization coefficient of the  hydrogen atom [22].}
\end{figure}
\begin{figure}[h]
\includegraphics[width=0.8\textwidth, height=220px]{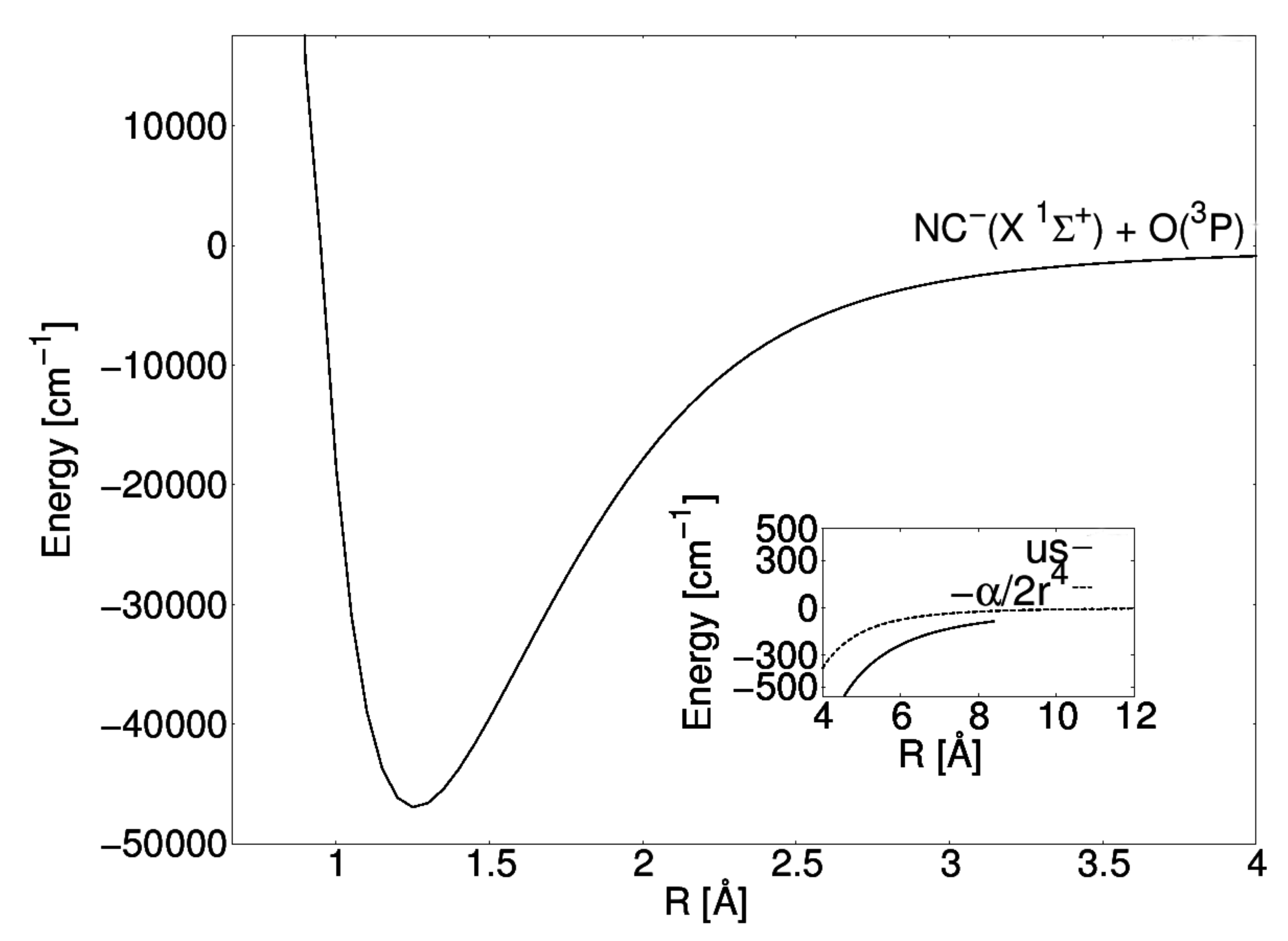}
\caption{The lowest electronic ground state of the $NCO^-$ system. In the smaller panel we compare the ab initio calculated points with the analytic extrapolation through the dipolar polarizability coefficient. Here $\alpha$ = 0.889 ${\buildrel _{\circ} \over {\mathrm{A}}}^3$ is the polarization coefficient of the oxygen atom [23].}
\end{figure}
Such data are providing a measure for the partners' localization within  the well. It is interesting to note here that from the calculated PEC of $\mathrm{C_2H^-}$ from reference [15] we found 21 bound levels along the additional bond between $\mathrm{C_2^-}$ and the hydrogen atom, while our present calculations  generate a  potential energy curve which supports 24 vibrational levels, as expected from  the increase in the well depth and the stronger tail of the long-range part of the PEC. 
If we turn now to the interaction in the $\mathrm{NCO^-}$ anion, we note that the full potential energy curves for its lowest-lying electron state $\mathrm{X ^1\Sigma^+}$ and for the next higher electronic excited states have been already presented elsewhere [19], where it was shown that the ground state of $\mathrm{NCO^-}$ is stable with respect to electron detachment processes and it is also energetically  well separated from the excited electronic states of both anion and neutral radical species. The one dimensional cut for the collinear arrangement of the pseudo-2Body potential  for the lowest electronic ground state PEC of $\mathrm{NCO^-}$, when considering only the newly formed bond with the incoming oxygen atom,is given in Figure 2 along the C-O distance. Also for this system the ab initio  points in the long range region have been compared with the points provided by an  analytic extrapolation. In this this case as an extrapolation coefficient we took the polarizability coefficient of the oxygen $\mathrm{O}$ atom  in its ground electronic state. The computed curve asymptotically refers to $\mathrm{NC^-}$ + $\mathrm{O}$ separate electronic energies in the states reported by eq.(3). The vibrational bound states which we have computed for this new bond in the  $\mathrm{NCO^-}$ system are given in Table 2. From the table we can see that the changes in masses and well depth in comparison with the previous case allow this  potential to accomodate up to 106 bound states for the zero angular momentum arrangement. The ZPE value for the complex is also reported.
It is important to note here that the ZPE values we are reporting refer to the single, 1D PECs associated with each of the two new bonds formed in each molecule.Hence the total 3-atom molecular complexes will have a different value in each case. The latter,however, is not directly relevant to the present study and calculations.
Another structural issue  in the present approach to the RA path calculations is that of the most likely geometries that could most efficiently preside over the final formation of the two closed-shell anions discussed in this work. In particular, keeping the stronger bonds of the anionic fragments fixed during the approach of either atomic partner suggests a sort of "adiabatic" picture to the final formation of the triatomic anions. Such a situation of a nearly non-rotating molecular anionic partner (i.e. either $\mathrm{C_2^-}$ or $\mathrm{NC^-}$)  would be in keeping with the expected cold environments provided by the DMC, an environment largely screened from the photon-dominated regions (PDR) that would instead reduce the  content of the initial anionic partners. In other words, it is a reasonable simplification to investigate the process of radiative stabilization as occurring at  relative approaching velocities whereby the atomic partners interact with rotationally "cold" anionic diatomics and therefore the relative angles play the role of an "adiabatic" parameters. In such a somewhat simplified view, one can then argue that it is important to further examine the full interactions for different "cuts" of the triatomic surface provided by different choices of the relative Jacobi angles, in order to establish an "efficiency order" of the RA paths as a function of the adiabatic angle.
 
\begin{center}
 \begin{tabular}{ | l | l | l | }
    \hline n & Present [$\mathrm{cm^{-1}}$] & $\mathrm{S/H^*}$ [$\mathrm{cm^{-1}}$] \\ \hline
    0  & -36484.712 & -35101.0503 \\
    1  & -33382.515 & -31740.7585 \\
    2  & -30400.696 & -28575.3909 \\
    3  & -27540.899 & -25581.6902 \\
    4  & -24805.149 & -22757.6764 \\
    5  & -22195.895 & -20109.0782 \\
    6  & -19716.045 & -17633.9233 \\
    7  & -17368.989 & -15334.5300 \\
    8  & -15158.555 & -13213.1755 \\
    9  & -13088.914 & -11264.2377 \\
    10 & -11164.350 & -9481.0847 \\
    11 & -9388.864 & -7855.3960 \\
    12 & -7765.712 & -6377.7348 \\
    13 & -6296.939 & -5042.9928 \\
    14 & -4983.233 & -3852.4982 \\
    15 & -3824.272 & -2810.9689 \\
    16 & -2819.461 & -1928.7935 \\
    17 & -1968.859 & -1220.8291 \\
    18 & -1273.799 & -690.6834 \\
    19 & -736.663 & -317.5733 \\
    20 & -358.695 & -93.9227 \\
    21 & -132.461 &  \\
    22 & -31.695 & \\
    23 & -4.527 & \\
    \hline
    \hline ZPE [$\mathrm{cm^{-1}}$] us & ZPE [$\mathrm{cm^{-1}}$] $\mathrm{S/H^*}$ \\ \hline
    1522.7532 & 1740.6574 \\ \hline
    \hline
    \end{tabular}
\captionof{table}{Bound vibrational levels (J = 0) for the new bond formed in the ground electronic state of $\mathrm{C_2H^-}$ using ourpreviously computed PECs.}
\end{center}

\begin{center}
    \begin{tabular}{ | l | l | l | l | l | l | l | l |}
    \hline $\mathrm{\nu}$ & E[$\mathrm{cm^{-1}}$] & $\mathrm{\nu}$ & E[$\mathrm{cm^{-1}}$] & $\mathrm{\nu}$ & E[$\mathrm{cm^{-1}}$] & $\mathrm{\nu}$ & E[$\mathrm{cm^{-1}}$]  \\ \hline
0 & -46844.0141 & 28 & -19387.9834 & 56 & -3972.5315 & 82 & -239.3811 \\ \hline
1 & -45643.1785 & 29 & -18626.7360 & 57 & -3659.5916 & 83 & -209.9141 \\ \hline
2 & -44460.5593 & 30 & -17880.3572 & 58 & -3363.3565 & 84 & -183.6564 \\ \hline
3 & -43295.8876 & 31 & -17148.9104 & 59 & -3083.7068 & 85 & -160.2811 \\ \hline
4 & -42148.8837 & 32 & -16432.4636 & 58 & -2820.4982 & 86 & -139.4903 \\ \hline
5 & -41019.3065 & 33 & -15731.0848 & 59 & -2573.5558 & 87 & -121.0300 \\ \hline
6 & -39906.9157 & 34 & -15044.8436 & 60 & -2342.6087 & 88 & -104.6750 \\ \hline
7 & -38811.3766 & 35 & -14373.8183 & 61 & -2127.1929 & 89 & -90.2136 \\ \hline
8 & -37732.3834 & 36 & -13718.1062 & 62 & -2342.6087 & 90 & -77.4599 \\ \hline
9 & -36669.7251 & 37 & -13077.8245 & 63 & -2127.1929 & 91 & -66.2346 \\ \hline
10 & -35623.2323 & 38 & -12453.0945 & 64 & -1926.9946 & 92 & -56.3825 \\ \hline
11 & -34592.7019 & 39 & -11844.0019 & 65 & -1741.6563 & 93 & -47.7554 \\ \hline
12 & -33577.9296 & 40 & -11250.5821 & 66 & -1570.6160 & 94 & -40.2207 \\ \hline
13 & -32578.7184 & 41 & -10672.9211 & 67 & -1413.4147 & 95 & -33.6573 \\ \hline
14 & -31594.9113 & 42 & -10111.2251 & 68 & -1269.3628 & 96 & -27.9566 \\ \hline
15 & -30626.3606 & 43 & -9565.6640 & 69 & -1137.8252 & 97 & -23.0204 \\ \hline
16 & -29672.9412 & 44 & -9036.2330 & 70 & -1018.1257 & 98 & -18.7620 \\ \hline
17 & -28734.5478 & 45 & -8522.9578 & 71 & -909.5335 & 99 & -15.1014 \\ \hline
18 & -27811.0882 & 46 & -8026.0183 & 72 & -811.3535 & 100 & -11.9677 \\ \hline
19 & -26902.4868 & 47 & -7545.5452 & 73 & -722.7938 & 101 & -9.2924 \\ \hline
20 & -26008.6812 & 48 & -7081.5849 & 74 & -642.9654 & 102 & -6.9957 \\ \hline
21 & -25129.6252 & 49 & -6634.2286 & 75 & -571.2271 & 103 & -4.9344 \\ \hline
22 & -24265.2855 & 50 & -6203.5597 & 76 & -506.7893 & 104 & -2.9930 \\ \hline
23 & -23415.6428 & 51 & -5789.6381 & 77 & -448.9800 & 105 &  -1.1570 \\ \hline
24 & -22580.6885 & 52 & -5392.5183 & 78 & -397.1795 \\ \hline
25 & -21760.4263 & 53 & -5012.2370 & 79 & -350.8530 \\ \hline
26 & -20954.8696 & 54 & -4648.8161 & 80 & -309.4392 \\ \hline
27 & -20164.0435 & 55 & -4302.2562 & 81 & -272.4099 \\ \hline

\hline
    \end{tabular}
\begin{tabular}{ | l | l | }
\hline ZPE [$\mathrm{cm^{-1}}$] & 605.6033 \\ \hline
\hline
\end{tabular}
\captionof{table}{Bound vibrational levels (J = 0) for the ground electronic state of the newly formed bond in the $\mathrm{NCO^-}$ system using the  1D PEC computed in the present work.}
\end{center}

In Figures 3 and 4 we have therefore reported the change of the PECs at different intermolecular angles for $\mathrm{C_2H^-}$ and $\mathrm{NCO^-}$ respectively. 
The data in Figure 3 for the $\mathrm{C_2H^-}$ system were  obtained at the MRCI level of calculation using the aug-cc-pVTZ basis set, while Figure 4 corresponds to $\mathrm{NCO^-}$  results which have been calculated at the MP2 level with aug-cc-pVQZ basis set. The given angles for two complexes are associated with the  coordinates of the triatomic system, where the stronger and stiffer  multiple  bond is kept fixed: the angle is different for each PEC shown and the H(O) distances are being also varied for each fixed angle. The angle is that formed by the main radial variable with respect to either the $\mathrm{C_2^-}$ or the  $\mathrm{NC^-}$ bonds. By looking at the locations of the well depths as a function of the chosen angles we can see that in both systems the PECs for the collinear case  are exhibiting the deepest wells, while the "bent" configurational approaches afford weaker interactions where that depth decreases by more than 30.% of its collinear value. This feature indicates that the complexes formed for this configuration will be the most  stable and will therefore provide a higher density of states for the final complex to get formed by emitting a photon.We shall see in the next section that the availability of more final rotovibrational states for the anionic complexes indeed causes larger RA cross sections and larger recombination rates, all other properties being equal. We therefore decided that it may be more significant for the present exploration to focus on those RA processes which would follow the collinear paths to recombination, as they would here represent the most favourable conditions. The calculations detailed below will further discuss this point.

\begin{figure}[b]
\includegraphics[width=0.9\textwidth, height=270px]{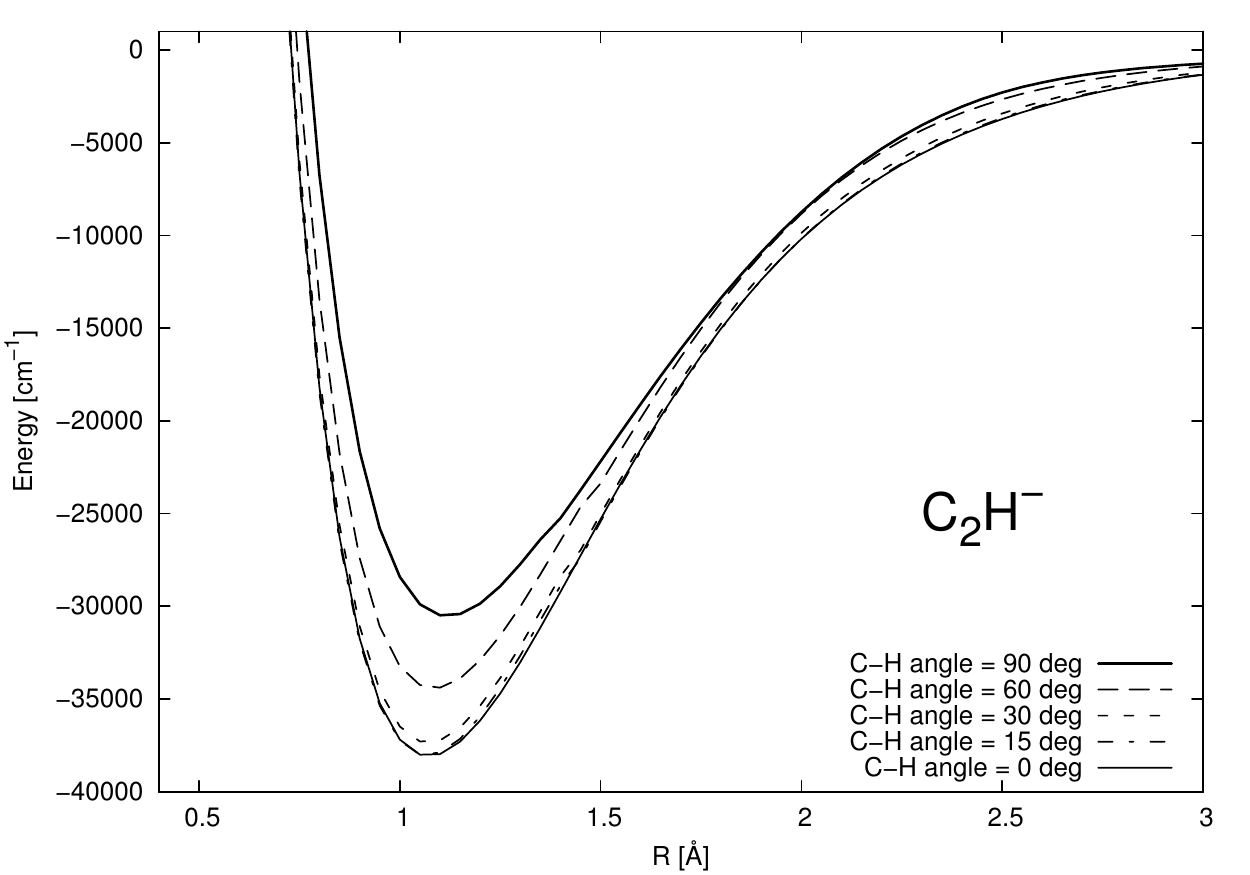}
\caption{The ground electronic PECs for the $\mathrm{C_2H^-}$ anionic complex computed for  different intermolecular angles. The distance represents the outer carbon's bond to the H atom. See main text for further details.}
\end{figure}

\begin{figure}[t]
\includegraphics[width=0.9\textwidth, height=270px]{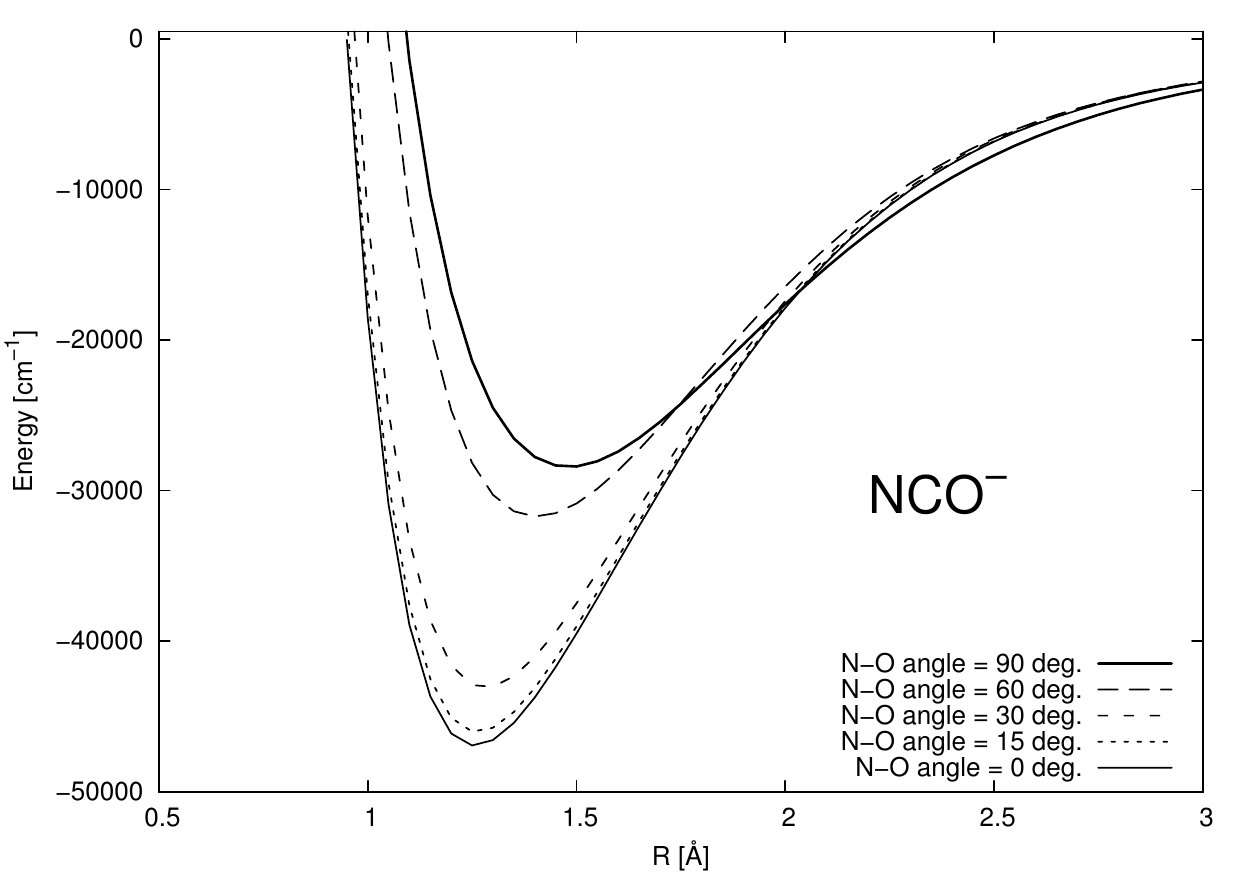}
\caption{The ground electronic PECs of the $\mathrm{NCO^-}$ system at different intermolecular angles.The distance represents the carbon's bond to the oxygen atom. See main text for further details.}
\end{figure}

\subsection{ELECTRIC DIPOLE MOMENTS OF $\mathrm{\mathbf{C_2H^-}}$ AND $\mathrm{\mathbf{NCO^-}}$}

The dipole functions associated with the pseudo-1D interaction potentials employed in this study are important properties, as will become clearer in the following section,  which need to be computed for the quantum treatment of the RA mechanisms involving  the present  systems . For both  species our computed $\mu(R)$ was obtained between 0.7 ${\buildrel _{\circ} \over {\mathrm{A}}}$ and 18 ${\buildrel _{\circ} \over {\mathrm{A}}}$ and the results of the  calculations are given in Figure 5. We can clearly see there that , within the examined  distance range, the overall radial behaviour of the dipole functions are qualitatively similar, although the one for the $\mathrm{NCO^-}$ system exhibits a much steeper dependence on that variable.The Figure further reveals an interesting behaviour for the complexes: when analysing the partial Mulliken charges on both $\mathrm{C_2^-}$ and $\mathrm{H}$ fragments of $\mathrm{C_2H^-}$ and the $\mathrm{NC^-}$ and $\mathrm{O}$ components of $\mathrm{NCO^-}$, we are able to see that, for example for $\mathrm{C_2H^-}$, at the smaller distances the negative charge sits on the $\mathrm{C_2^-}$ and therefore the dipole is positive and not small. However, when the distance increases  we see charge transfer effects from the $\mathrm{C_2^-}$ to the $\mathrm{H}$ that change the sign of the dipole which now goes down to zero, thereby indicating that most of the negative charge has  neutralized the $\mathrm{H}$ atom. 

\clearpage
\begin{figure}[t]
\includegraphics[width=0.9\textwidth, height=270px]{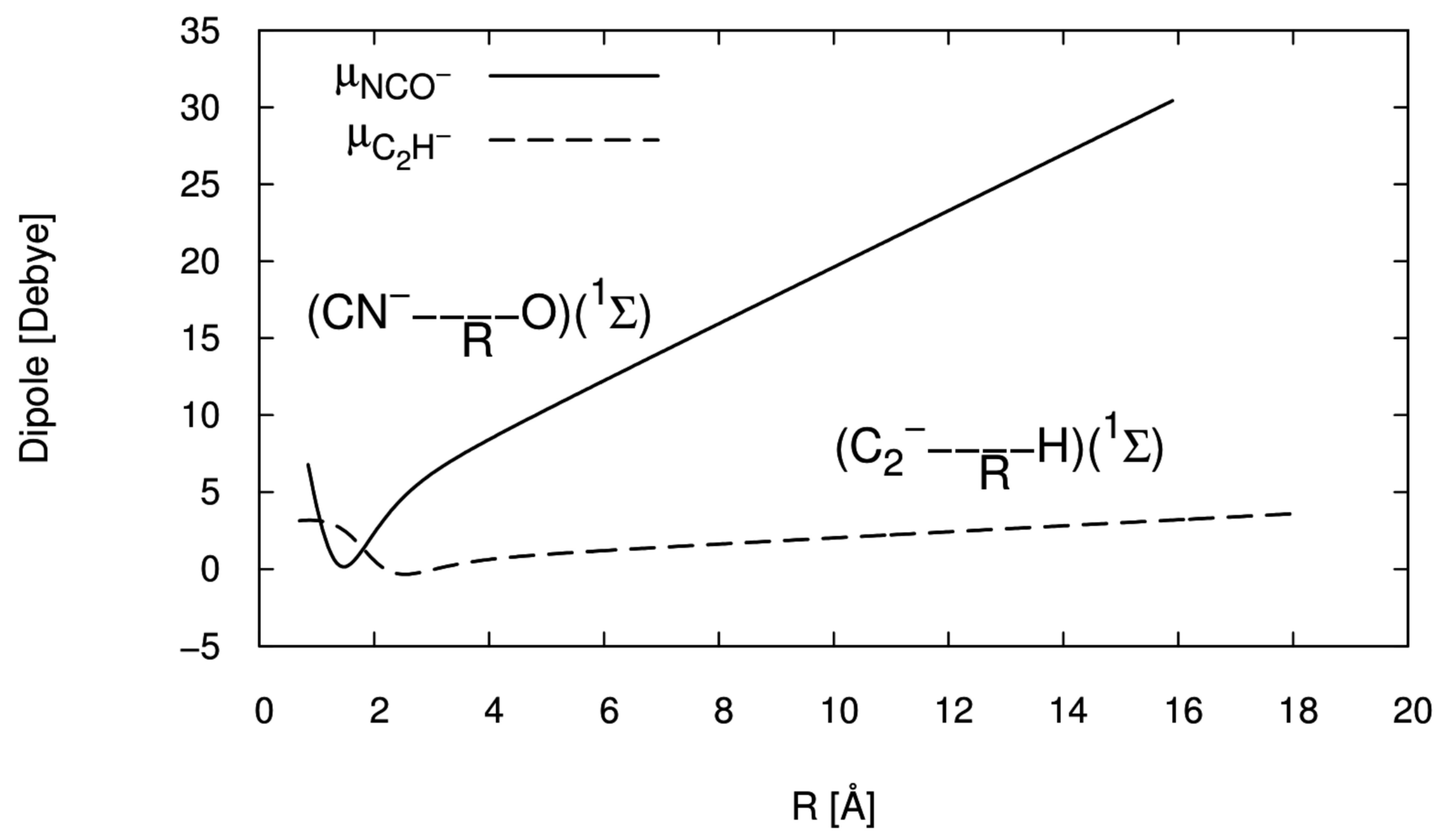}
\caption{Computed dipole moments for the ground electronic states of the two complexes as a function of the same internuclear distances already defined in Fig.s 3 and 4 and used here for the collinear approaches.
 See main text for further details.}
\end{figure}
As the relative distance is increased, we see that the negative charge, after a distance of  about 2 ${\buildrel _{\circ} \over {\mathrm{A}}}$, returns on the $\mathrm{C_2}$: the dipole therefore  increases with the  distance  while the opposite charges remain separate on the two fragments.The process is largely similar for the case of the $\mathrm{NCO^-}$, where however the effects are much more dramatic howing to the larger number of electrons involved in the biding and the greater polarization effects as charges are separated with the increasing of the relative distance. Hence, we are able to note from the calculations that at relatively large distances the increase in the value of the dipole function for the $\mathrm{NCO^-}$ complex is about one order of magnitude larger than in the case of the $\mathrm{C_2H^-}$ complex. As we shall show below, such structural findings are going to clearly affect the relative probabilities for forming the closed-shell anionic complexes by following the RA paths. As a final observation, it is interesting to note that similar calculations for the behaviour of the dipolar functions at different orientation angles follow largely the same patterns as those shown in figure 5, this being indeed the case for both systems. We have, in fact, computed the behaviour of the dipolar radial functions  by also changing the approaches of the H/O atoms from the those of the collinear configurations reported by that figure. We found in both cases that the general behaviour remains very close to those for the collinear configurations, the only difference being that their values at the shorther distances around 2.0 ${\buildrel _{\circ} \over {\mathrm{A}}}$ become smaller as the bending angles depart from linearity. One can therefore conclude from these numerical investigations that the collinear approach to RA processes remains  the most efficient for the RA dynamics, hence the one for which we shall carry out our calculations.

%\begin{figure}[t]
%\includegraphics[width=0.9\textwidth, height=270px]{NCO_C2H_anion_dipole.pdf}
%\caption{Computed dipole moments for the ground electronic states of the two complexes as a function of the same internuclear distances already defined in Fig.s 3 and 4 and used here for the collinear approaches.
% See main text for further details.}
%\end{figure}

\section{DYNAMICS OF PHOTOASSOCIATION PROCESSES}
\subsection{THE QUANTUM FORMULATION OF RA PROCESSES}

In the case of the radiative association mechanism, we shall follow the direct and resonant (indirect) pseudo-2Body  mechanism that will  become possible for the present systems once we select a fixed value of the orientational angle during the collisional interactions, a partial simplification of the dynamics which we have already discussed in the previous Section. More specifically, all calculations will be carried out following the collinear path for the approaching partners since such an arrangement shall maximise the efficiency of the RA processes. The stronger multiple bonds of the diatomic partners will also be kept constant during the calculations.
 
At a fixed, relative collision energy of the partners the Einstein coefficient for spontaneous emission constitutes a starting point in obtaining the cross sections. In atomic units such a coefficient is defined as [23,24]:
\begin{equation}
A_{vJ,v'J'} = \frac{32}{3c^3}\pi^3\nu_{vJ,v'J'}^3\frac{S_{JJ'}}{2J+1}M_{vJ,v'J'}
\end{equation}
Here $\mathrm{M_{vJ,v'J'}}$ is the transition dipole moment between the relevant rotovibrational states and it can be defined as following [25]:
\begin{equation}
M_{vJ,v'J'} = \int_{0}^{\infty}\phi_{vJ}(R)\mu(R)\phi_{v'J'}(R)dR
\end{equation}
and $\mathrm{S_{JJ'}}$ is the H{\"o}nl-London factor [24]. The required wavefunction for the final bound
state reached during the process can be  replaced by the initial wavefunction in the continuum  the preselected relative energy of the encounter. 
Then the transition moment is given by [26,27]:
\begin{equation}
M_{EJ,v'J'} = \int_{0}^{\infty}f_{EJ}(R)\mu(R)\phi_{v'J'}(R)dR
\end{equation}

The total cross section for a specific collision energy is also defined as [25,26]:
\begin{equation}
\sigma(E)=\frac{\pi}{k^2}\sum_J(2J+1)P(J)=\sum\sigma_K(E)
\end{equation}
 here $\mathrm{P(J)}$ is the probability for a given partial-wave component $\mathrm{f_{EJ}}$ to decay by photon emission. The terms of the summation define the contributions of the partial cross sections to the final process.
 In atomic units the partial cross section is given by:
\begin{equation}
\sigma_{J,v'J'}=\frac{64}{3}\frac{\pi^5}{c^3}\frac{p}{k^2}\nu_{Ev'J'}^3S_{JJ'}M_{EJ,v'J'}^2
\end{equation}
where $k^2=2\mu E$, with $\mathrm{E}$ being the collision energy, $\mathrm{\mu}$ is reduced mass, p the statistical weight of the initial electronic state. 
The evaluation of the total cross section for the case of the radiatively stimulated process is further given by the following expression:
\begin{equation}
\sigma_{st}=\frac{8}{3}\frac{\pi^4}{c}\frac{p}{k^2}\sum_{J,v',J'}I(\nu)S_{JJ'}M_{EJ,v'J'}^2
\end{equation}
where $\mathrm{I(\nu)}$ is the black body radiation field (BRF) [27], a quantity further characterized by the bath temperature $\mathrm{T_b}$
\begin{equation}
I(\nu)=\frac{4h\nu}{c^2}\frac{1}{exp(h\nu/kT_b)-1}
\end{equation}

here $h\nu=E-\epsilon_{v'J'}$ is emitted photon energy. The global, total cross section including both spontaneous and stimulated processes is finally given by [28]:
$$
\sigma(E)=\frac{64}{3}\frac{\pi^5}{c^3}\frac{p}{k^2}\sum_{J,v',J'}\nu_{Ev'J'}^3S_{JJ'}M_{EJ,v'J'}^2$$
\begin{equation}
\times\frac{1}{1-exp(-h\nu/kT_b)}
\end{equation}

The next step of the quantum analysis is that of  obtaining the corresponding total rates as a function of the environmental temperature [29]. It will require a further integration  over a Maxwellian distribution of the partner's relative velocities. This choice implies the modeling of the physical environment in the molecular clouds as being described via a local thermal equilibrium (LTE) condition:

\begin{equation}
K(T)=\left (\frac{8}{\mu\pi}  \right )^{1/2}\left ( \frac{1}{k_BT} \right )^{3/2}\int_0^\infty E\sigma(E)e^{-E/k_BT}dE
\end{equation}

\subsection{THE COMPUTED CROSS SECTIONS FOR THE RA PATHS}

The total cross sections  of the spontaneous and stimulated processes for the formation of either $\mathrm{C_2H^-}$ or $\mathrm{NCO^-}$ as a function of collision energy are given in figures 6, 7. The black body temperature for the photon bath in which the two species are immersed  ranges from 0 K up to 5000 K. 

\begin{figure}[b]
\includegraphics[width=0.9\textwidth, height=280px]{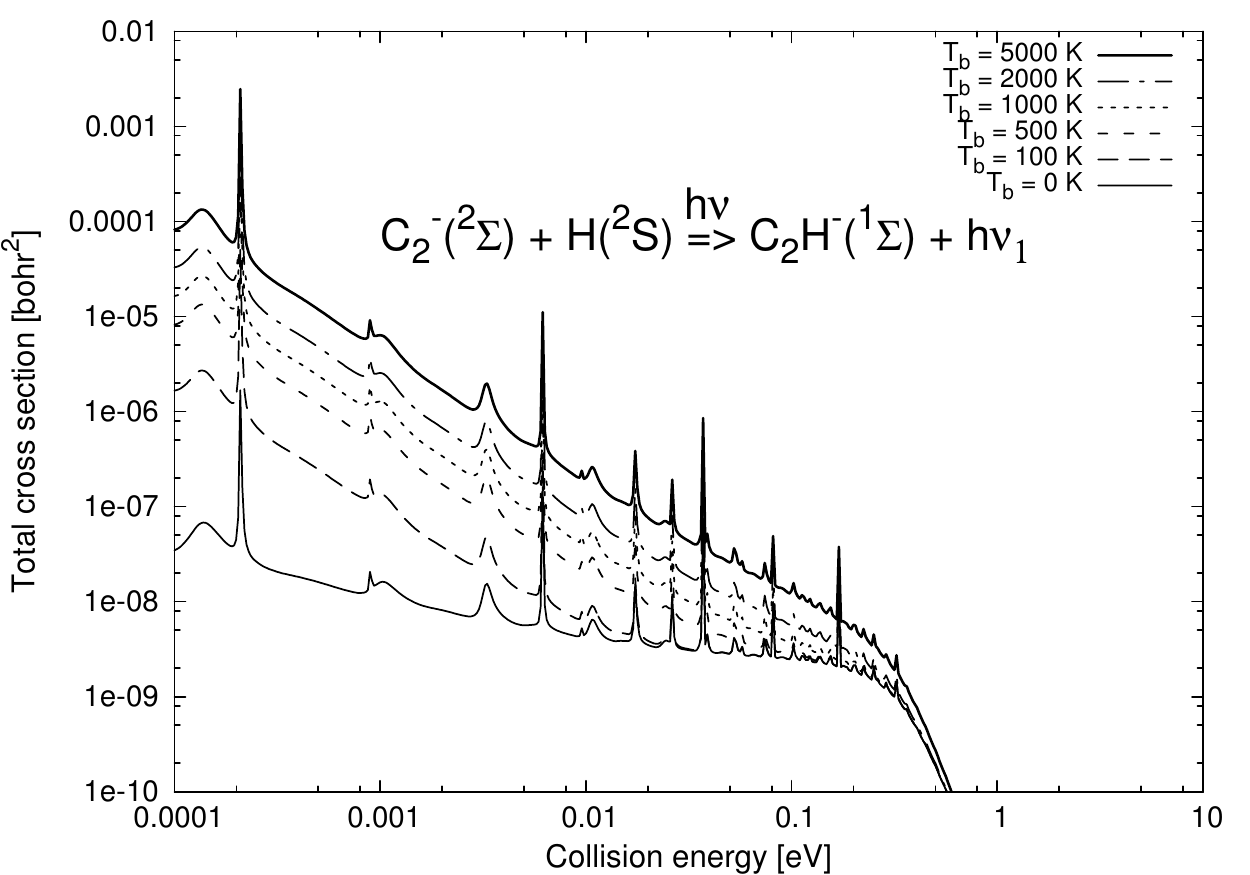}
\caption{Computed cross sections for stimulated plus spontaneous radiative association of $\mathrm{C_2^-}$ and $\mathrm{H}$ for various radiation temperatures $\mathrm{T_b}$ as a function of collision energy. The spontaneous associative mechanism is given by the curve labelled "0 K".}
\end{figure}

  The spontaneous process corresponds to $\mathrm{T_b=0}$ K. It can be seen from the figures that the presence of the BRF gradually enhances the magnitude of the cross sections. We can see that in the range of the smaller collision energies the stimulated process changes as a function of BRF temperature by three to four orders of magnitude the size of  the total association cross sections for $\mathrm{C_2H^-}$. For $\mathrm{NCO^-}$ the size of the change is even over five orders of magnitude. We should also note from the figures that the effect of the background radiation becomes however less important as one considers the largest collision energies: for increasingly faster approaching partners the RA cross sections uniformely decrease to nearly zero for all  values of  $\mathrm{T_b}$. The pronounced sharp peaks in the figures are due to the shape resonances, features that we shall further discuss below in more detail. They obviously occur because of the presence of  quasi-bound states of the complex in the continuum: as we shall see below, each contribution to the resonances comes from a specific quasi bound roto-vibrational state trapped by a specific  centrifugal barrier. Furthermore, if we compare the magnitude of the cross sections between the two species of this study, we see that the total cross sections for $\mathrm{NCO^-}$  formation are one order of magnitude larger. 

\begin{figure}[t]
\includegraphics[width=0.9\textwidth, height=280px]{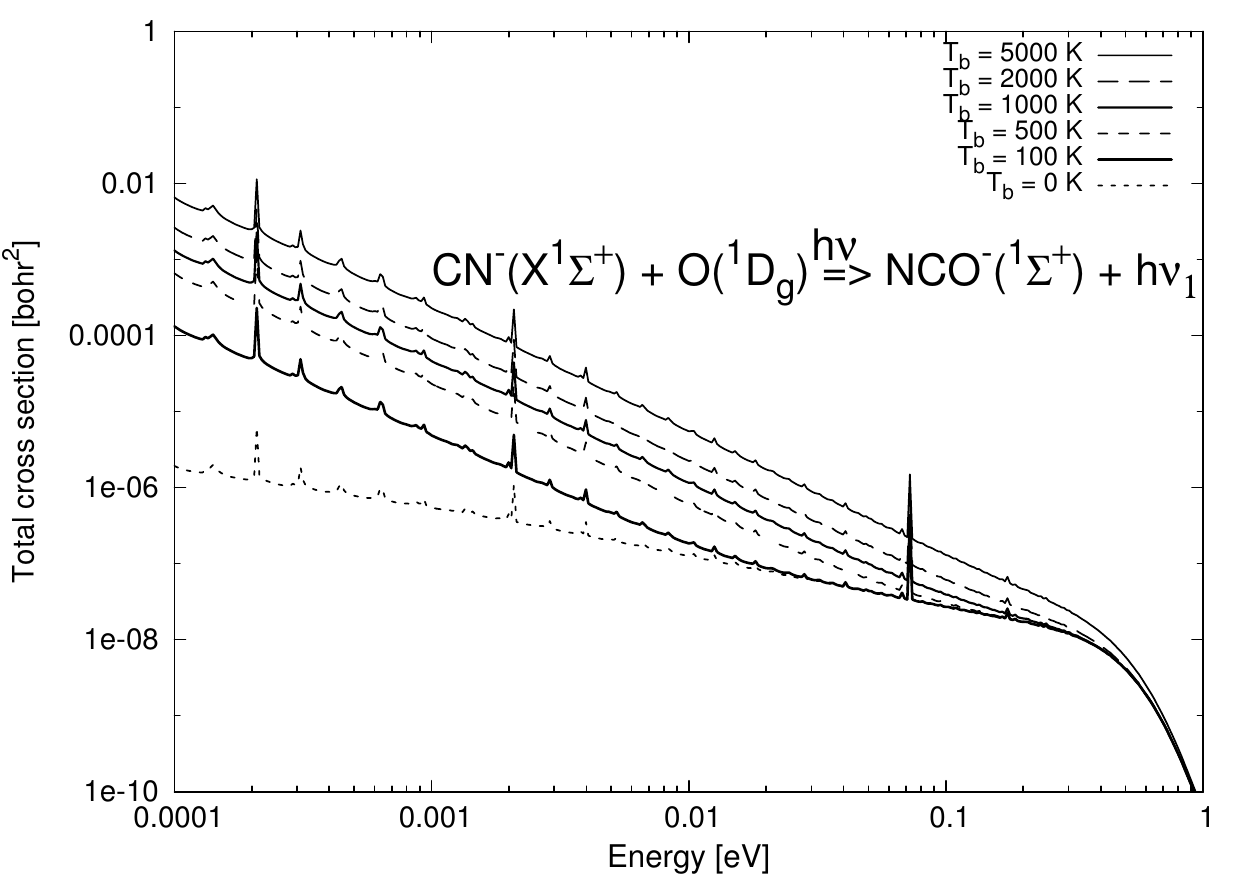}
\caption{Computed cross sections for stimulated plus spontaneous radiative association for $\mathrm{NC^-}$ and $\mathrm{O}$ for various radiation temperatures $\mathrm{T_b}$ as a function of collision energy. The spontaneous process corresponds to the curve labelled "0 K".}
\end{figure}

  The cross sections for the two species show qualitatively similar energy dependence since  they all decrease as the collision energy increases. This is due to the fact that, as the interaction time decreases, the time interval during which the photon can be emitted becomes shorter,thereby reducing the probability of the RA process to occur during the collisional event.

It is instructive to further look at the dependence of the partial cross sections on the number of bound vibrational states of the formed new bond at various bath temperatures for both $\mathrm{C_2H^-}$ and $\mathrm{NCO^-}$: the results of the present calculations are given by figures 8, 9. 
For both cases the value of computed $\mathrm{\sigma_{\nu}(E)}$ has been obtained by fixing the collision energy at the first resonant peak. It is interesting to note that the largest partial cross sections appear at $\mathrm{\nu}$ = 16 for $\mathrm{C_2H^-}$ while the other partial contributions decrease in value as $\mathrm{\nu}$ increases up to 23. For $\mathrm{NCO^-}$, the overall behaviour is similar: the largest value of the partial cross section occurs at $\mathrm{\nu}$ = 51 while the sizes steadily decrease as the vibrational final state increases up to 105. We can conclude from the behavior of both systems that the RA processes give rise to vibrationally excited products with probabilities of formation that decrease by several orders of magnitude as the final bound complex is formed in the lowest vibrational states. Naturally, all such bound anionic complexes  can, in principle, be detected by a following spontaneous emission of rotovibrational photons as they decay to the lower levels of the bound system. 

The uniform  increase of the partial cross sections up to a certain vibrational level for a given relative energy of approach between partners could be understood if one considers the overlap of
 the complex's wavefunctions in the continuum and those  of the final bound, and variously excited, vibrational states. For instance, when $\mathrm{\nu}$ = 16 for the case of $\mathrm{C_2H^-}$, then the overlap reaches its maximum. The decrease of the $\mathrm{\sigma_{\nu}(E)}$ as one moves either to more deeply bound final complexes (lower vibrational quantum numbers)  or to the bound complexes associated to highly excited vibrational levels  beyond the specific maximum in the cross sections, could be explained if we take into account both the value of the overlap integrals between molecular wavefunctions and the actual energy gap associated with the emitted photon. In particular, by looking at  eq.s (9), (10) and (12) where the role of the collison energy and the frequency of the emitted photon are clearly displayed.
\begin{figure}[b]
\includegraphics[width=0.9\textwidth, height=250px]{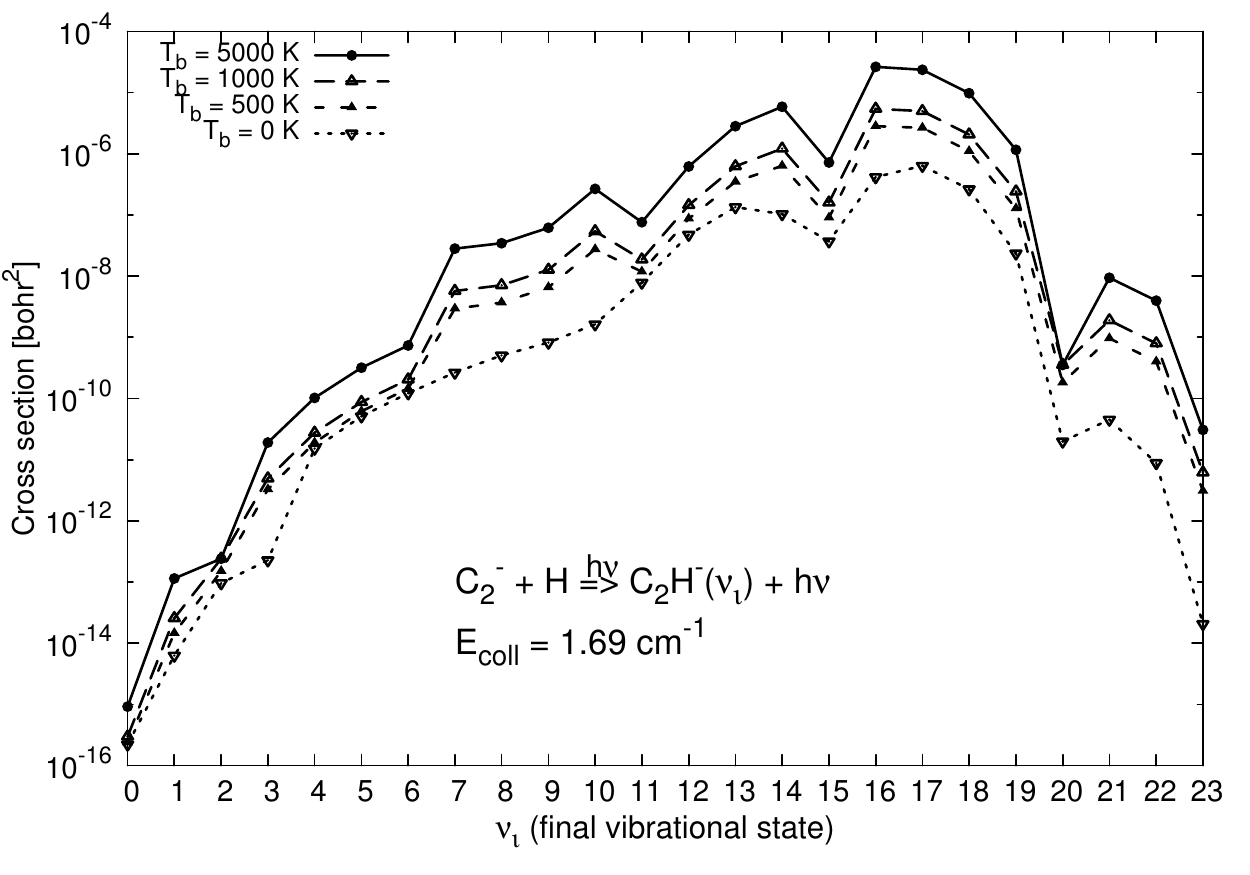}
\caption{Partial radiative asoociation cross sections as a function of the final vibrational level of the bound anionic complex  for $\mathrm{C_2H^-}$.The relative collision energy has been fixed as indicated in the Figure.}
\end{figure}
\clearpage
\begin{figure}[t]
\includegraphics[width=0.9\textwidth, height=250px]{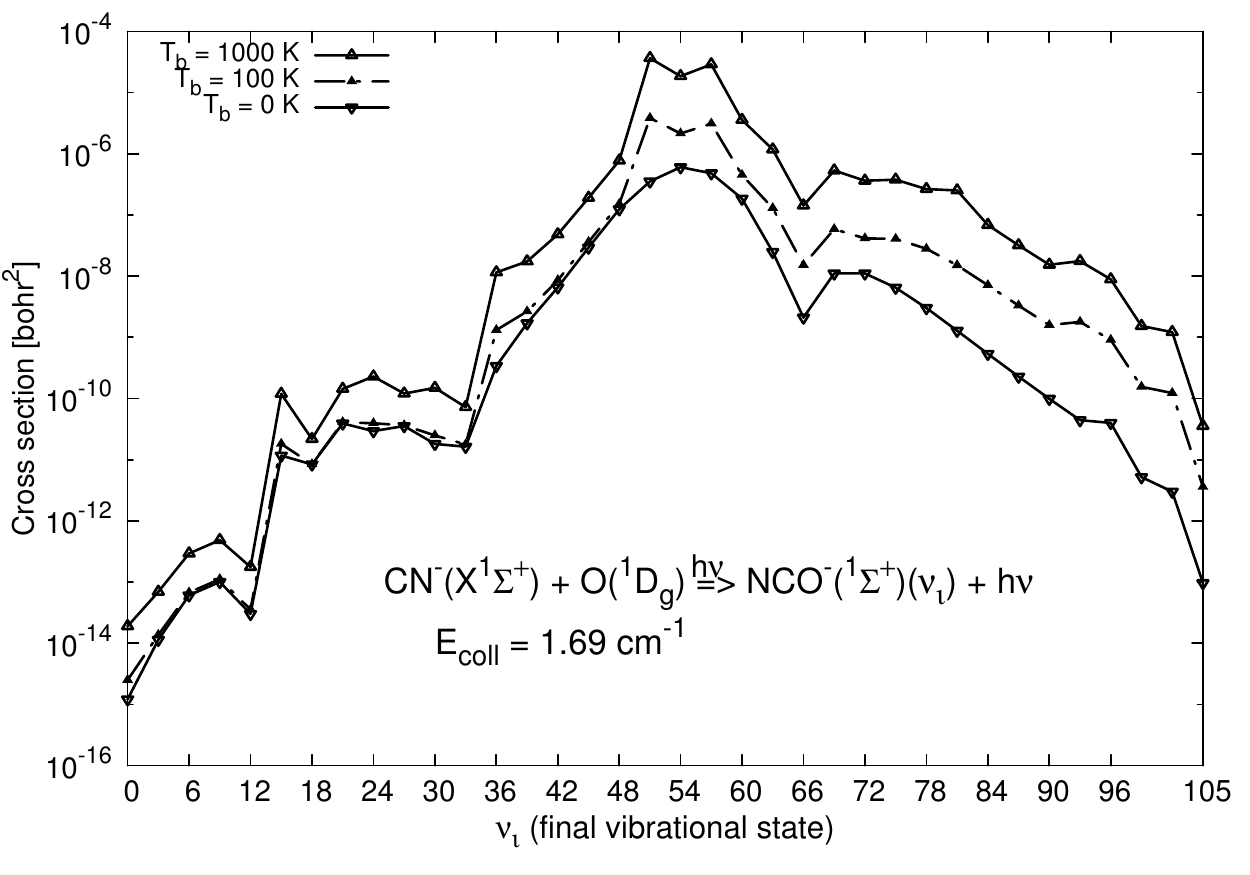}
\caption{Partial radiative asoociation cross sections as a function of final vibrational level of the bound anionic complex for $\mathrm{NCO^-}$ . The relative collision energy has been fixed at the value shown in the Figure.}
\end{figure}
%\begin{figure}[H]
%\includegraphics[width=0.9\textwidth, height=250px]{C2H_anion_xsecs_vib_state.pdf}
%\caption{Partial radiative asoociation cross sections as a function of the final vibrational level of the bound anionic complex  for $\mathrm{C_2H^-}$.The relative collision energy has been fixed as indicated in the Figure.}
%\end{figure}
%\begin{figure}[H]
%\includegraphics[width=0.9\textwidth, height=250px]{NCO_anion_xsecs_vib_state.pdf}
%\caption{Partial radiative asoociation cross sections as a function of final vibrational level of the bound anionic complex for $\mathrm{NCO^-}$ . The relative collision energy has been fixed at the value shown in the Figure.}
%\end{figure}

%\begin{figure}[H]
%\includegraphics[width=0.9\textwidth, height=260px]{C2H-_vib_number_resonance.pdf}
%\caption{Partial RA cross sections, as a function of the final vibrational levels of the bound anion, for the case of $\mathrm{C_2H^-}$}
%\end{figure}

%\begin{figure}[H]
%\includegraphics[width=0.9\textwidth, height=260px]{cross_section_vib_number_coll.pdf}
%\caption{Partial radiative association cross sections, as a function of the final vibrational levels of the bound molecular anion , for the case of $\mathrm{NCO^-}$}
%\end{figure}

\subsection{THE STRUCTURE OF THE RESONANT CHANNELS FOR THE PRESENT PROCESSES}

As  stated above, the pronounced structures that appear in figures 6, 7 are the result of shape resonances associated with the  metastable complex formations. In this section we will analyse in more detail the role and features of the resonances for both systems. At a selected positive energy, and for a given $\mathrm{J}$ value, the potential may support a metastable state located at an energy coinciding with that collision energy. Its wavefunction is therefore very similar, within the range of the potential, to that of a vibrational bound state and would therefore have larger transition moments to any of the lower-lying  bound vibrational states: hence the resonant enhancement of the corresponding cross section associated to that process.

In our case, the practical search of the resonance is linked to finding first a possible metastable state from any of the true bound states. They will appear at different energies and for different $\mathrm{J}$ values. Such a state will provide at a chosen energy  the initial state , while the final states will be given by various different bound vibrational states: the more of them involved, the stronger the resonance effects on the final cross sections.

 The first search is therefore that of producing the partial cross sections at the specific resonant energies, selected from the total cross sections as a function of the vibrational states: 

%\begin{figure}[H]
%\includegraphics[width=0.9\textwidth, height=260px]{C2H-_vib_number_resonance.pdf}
%\caption{Partial RA cross sections, as a function of the final vibrational levels of the bound anion, for the case of $\mathrm{C_2H^-}$}
%\end{figure}

%\begin{figure}[H]
%\includegraphics[width=0.9\textwidth, height=260px]{cross_section_vib_number_coll.pdf}
%\caption{Partial radiative association cross sections, as a function of the final vibrational levels of the bound molecular anion , for the case of $\mathrm{NCO^-}$}
%\end{figure}
the results from our present calculations are given by figures 10 and 11 and were discussed before. For both the $\mathrm{C_2H^-}$ and $\mathrm{NCO^-}$ complex formation we  have then chosen three sharp peaks corresponding to specific  resonant collision energies and further  selected the top three partial cross sections for specific final vibrational levels, plotting them as a function of energy around the region of the  pre-selected resonant peaks in figures 10, 11. From figure 12 we can see that the resonances persist for all selected vibrational states. Thus, for example in case of $\mathrm{C_2H^-}$, the resonance at a collision energy of 1.69 $\mathrm{cm^{-1}}$ is initiated by $\mathrm{\nu}$ = 22 and $\mathrm{J}$ = 6 but ends up in different final vibrational states. Other resonances which appear higher up in energy will come from other $\mathrm{\nu}$ values (usually lower values and associated with a larger $\mathrm{J}$) and will also show up in the partial cross sections of specific final states. It is the variety of such resonance processes that will cause the final cross sections to increase over different ranges of relative collision energies. Such changes in size shall obviously affect the ensuing association rates that we shall discuss in the following subsection.

\begin{figure}[b]
\includegraphics[width=0.9\textwidth, height=260px]{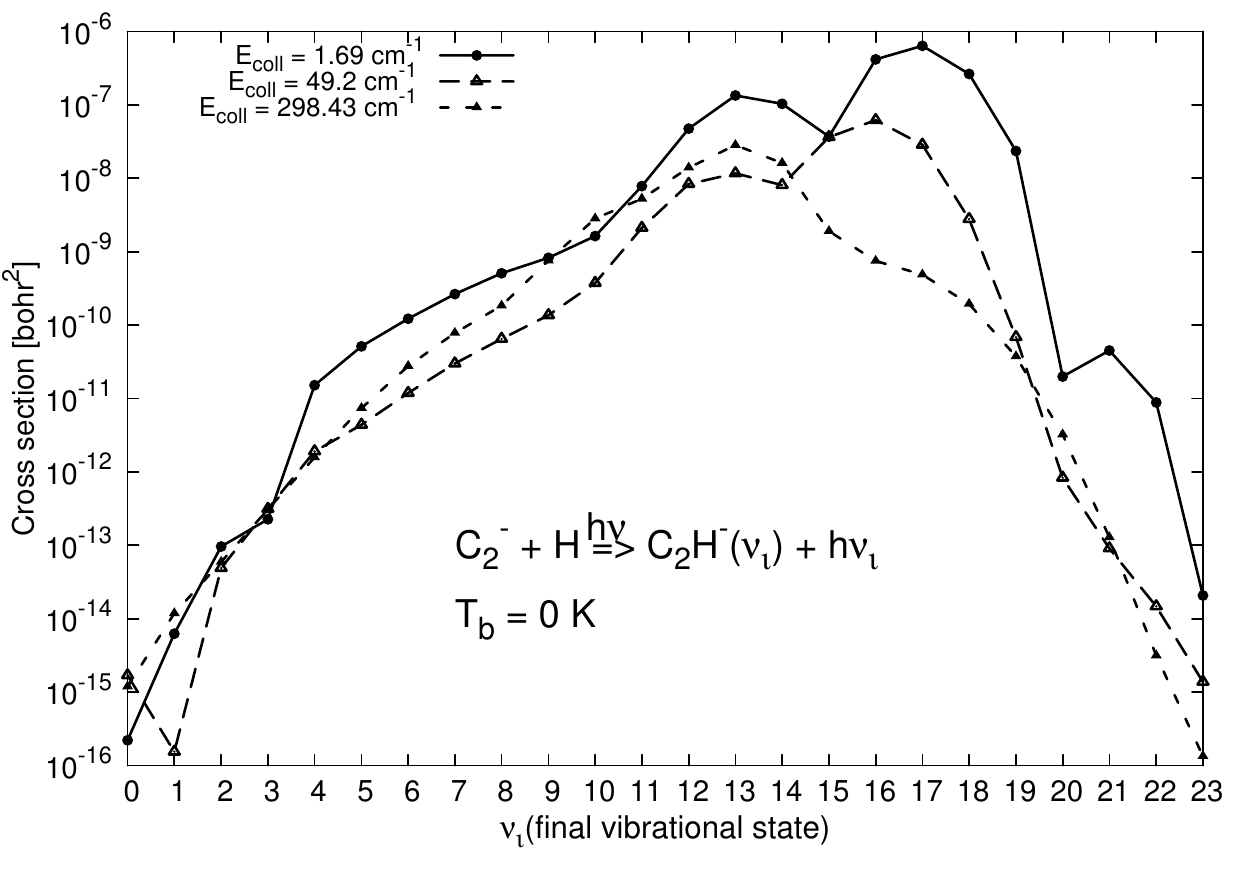}
\caption{Partial RA cross sections, as a function of the final vibrational levels of the bound anion, for the case of $\mathrm{C_2H^-}$}
\end{figure}
\begin{figure}[t]
\includegraphics[width=0.9\textwidth, height=260px]{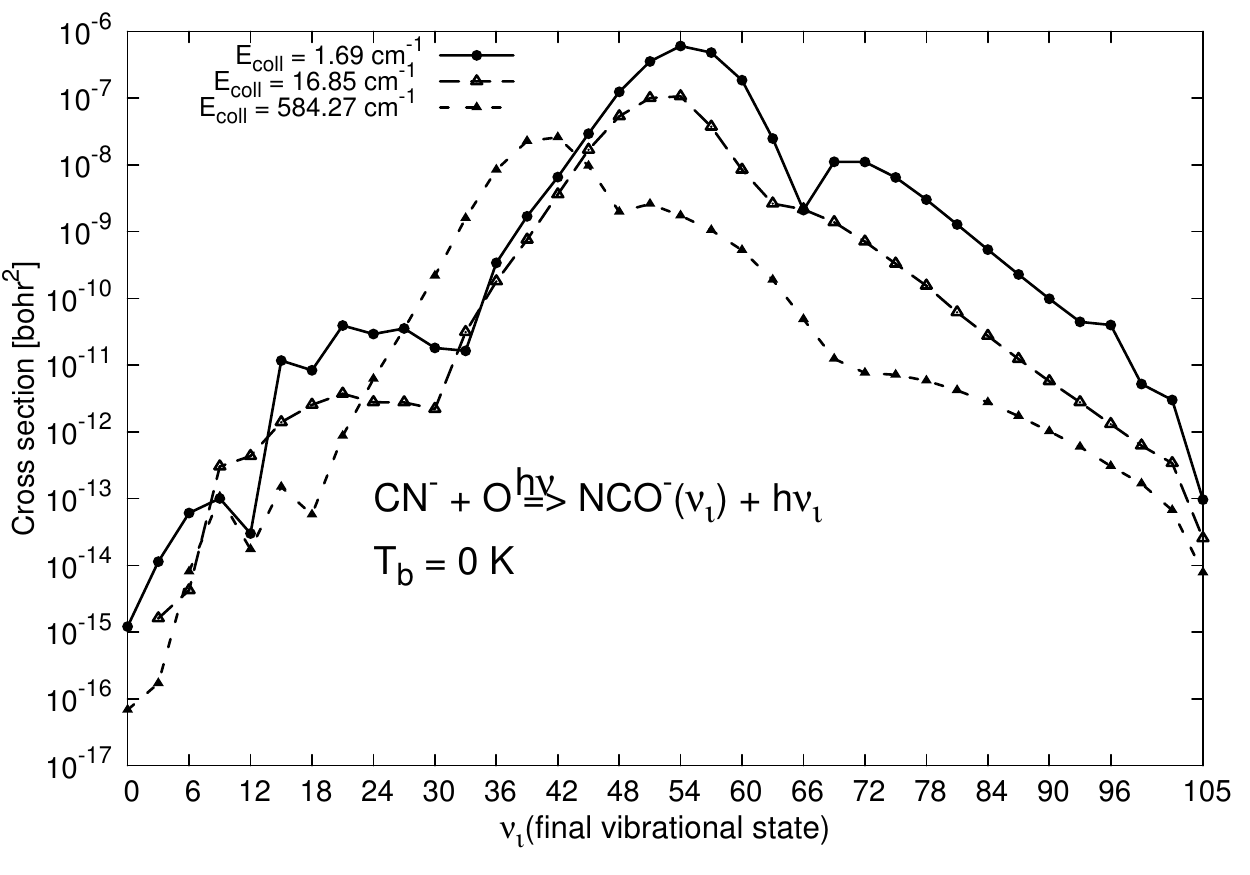}
\caption{Partial radiative association cross sections, as a function of the final vibrational levels of the bound molecular anion , for the case of $\mathrm{NCO^-}$}
\end{figure}

\clearpage
\begin{figure}[t!]
\includegraphics[width=0.9\textwidth, angle=-90]{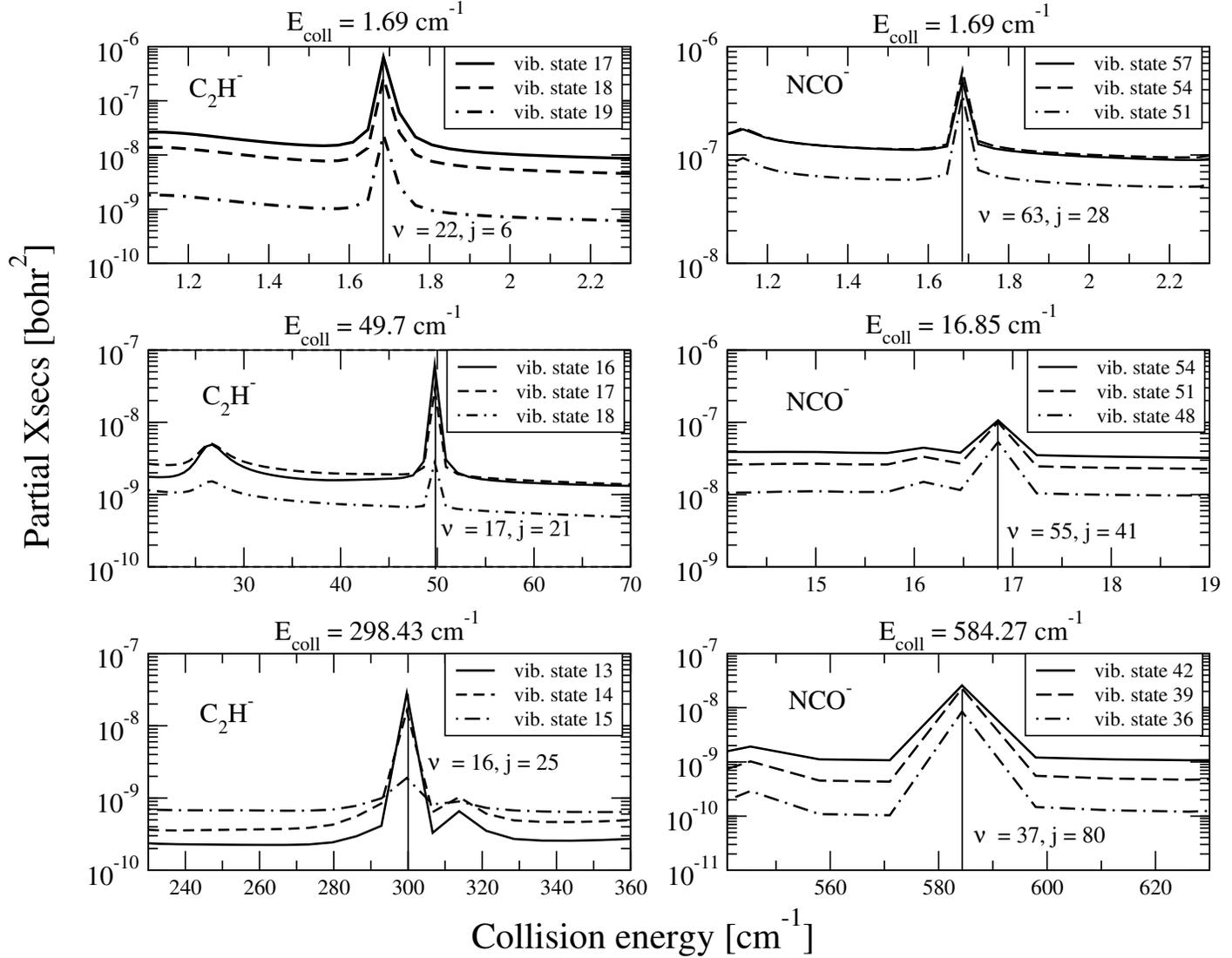}
%\subfigure{
%  \includegraphics[width=0.47\textwidth]{C2H-_1st_resonance.pdf}}
%\quad
%\subfigure{
%  \includegraphics[width=0.47\textwidth]{NCO-_1st_resonance.pdf}}
%
%\medskip
%\subfigure{
%  \includegraphics[width=0.47\textwidth]{C2H-_2nd_resonance.pdf}}
%\quad
%\subfigure{
%  \includegraphics[width=0.47\textwidth]{NCO-_2nd_resonance.pdf}}
%
%\medskip
%\subfigure{
%  \includegraphics[width=0.47\textwidth]{C2H-_3rd_resonance.pdf}}
%\quad
%\subfigure{
%  \includegraphics[width=0.47\textwidth]{NCO-_3rd_resonance.pdf}}
\caption{Shape resonances linked to different  metastable roto-vibrational states for $\mathrm{C_2H^-}$ and $\mathrm{NCO^-}$ at fixed collision energies.See main text for further details}
\end{figure}

\subsection{THE PARTIAL AND TOTAL RA RATES FOR $\mathrm{C_2H^-}$ and $\mathrm{NCO^-}$ FORMATION}

Rate coefficients for the complexes have been computed by numerical quadrature of Eq. (13) for a set of gas temperatures $\mathrm{T_g}$ ranging from 1 to 10,000 K. In the present  calculations the rates for spontaneous as well as for stimulated processes are reported in figures 13 and 14 for the two molecular anions. The total rates of formation for the spontaneous process exhibit a slow increase with the lower temperatures $\mathrm{T_g}$ and a fast decrease with $\mathrm{T_g}$ as the temperature goes above 1000 K. In the temperature region from 10 to 100 K the spontaneous RA rates for the $\mathrm{C_2H^-}$ anion formation  vary in size between $\mathrm{10^{-21}}$ $\mathrm{cm^3/s}$  and $\mathrm{10^{-20}}$ $\mathrm{cm^3/s}$, while in the case of the formation of the  $\mathrm{NCO^-}$ anion the corresponding rates vary between $\mathrm{10^{-19}}$ $\mathrm{cm^3/s}$  and $\mathrm{10^{-18}}$ $\mathrm{cm^3/s}$, in line with what should be expected from their difference in the density of final vibrational states available.
 
 In order to better understand how large  our rates are within the context of similar RA processes that can occur in DMC environments  for other molecular systems, we  compare them  below with three molecular cations for which radiative association  calculations already exist: $\mathrm{LiH^+}$ [28], $\mathrm{HeH^+}$ [29] and $\mathrm{CH^+}$ [30]. Their results are reported in the references indicated for each of them. It is important to note here that  for these systems the rates have been computed using fairly accurate ab initio PECs and the processes involve a single potential as in our calculations. The comparison at a sampling temperature of 50 K is given by the data of Table 3.
It is interesting to see there that the present results for both anions are very different from each other, those for $\mathrm{NCO^-}$ being about two orders of magnitude larger than those for $\mathrm{C_2H^-}$. Furthermore, the RA rates at the same temperature for the formation of two common cations in the modeling of the Early Universe evolution turn out to be either much smaller or of similar size, but not larger, than those of our present anions. Additionally, the $\mathrm{CH^+}$ cation is usually considered the first step towards the creation of larger hydrocarbons and it is the first molecular ion discovered in the interstellar medium [31,32]: its RA rate turns out to be markedly larger by various orders of magnitude than those of the present anions. It indicates the broad range of association efficiencies that simple molecular ions can exhibit under the cold environments of the ISM in the DMC chemistry.

\begin{figure}[h]
\includegraphics[width=0.9\textwidth, height=240px]{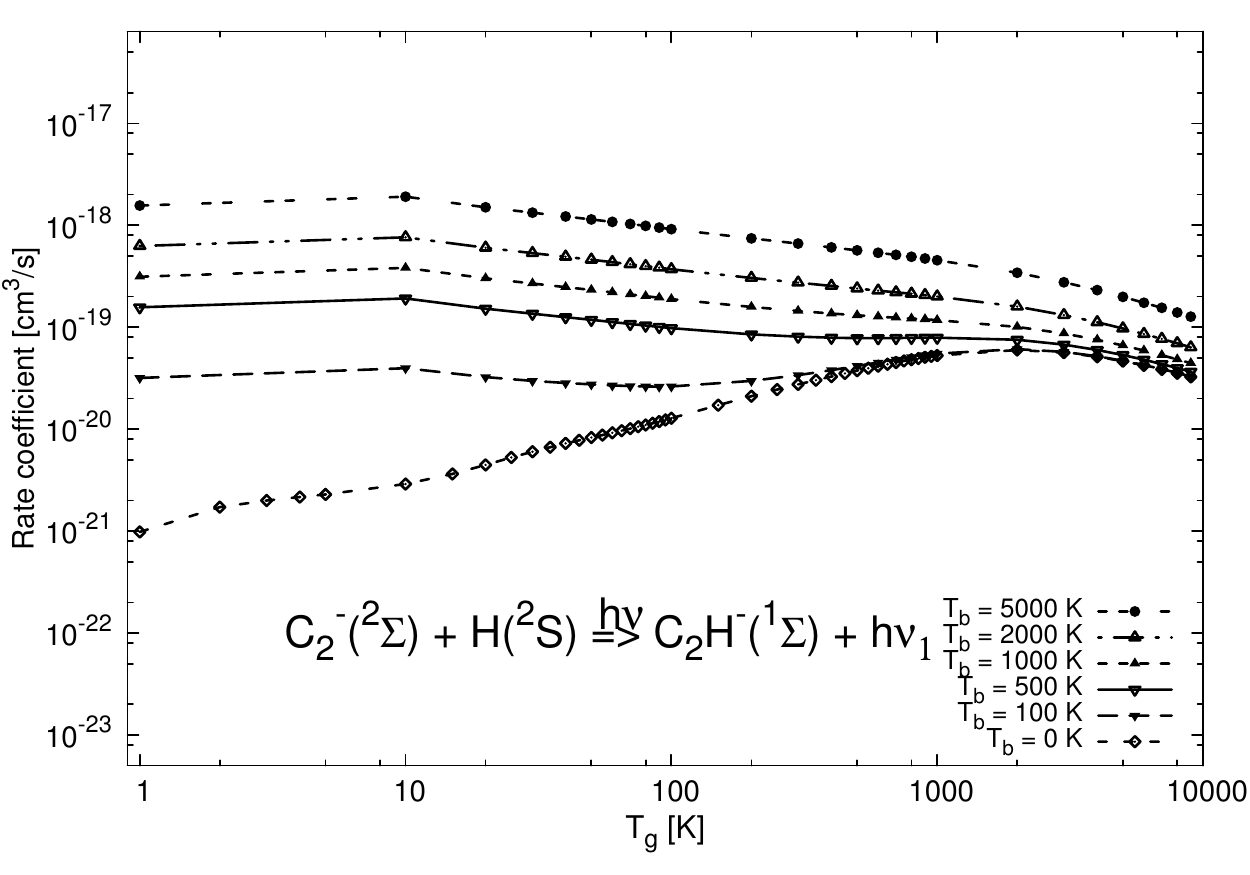}
\caption{Rate coefficients for stimulated and spontaneous radiative association of $\mathrm{C_2^-}$ and $\mathrm{H}$ for various blackbody radiation temperatures. The spontaneous process is associated with the curve labelled "0 K".}
\end{figure}

\begin{figure}[h]
\includegraphics[width=0.9\textwidth, height=240px]{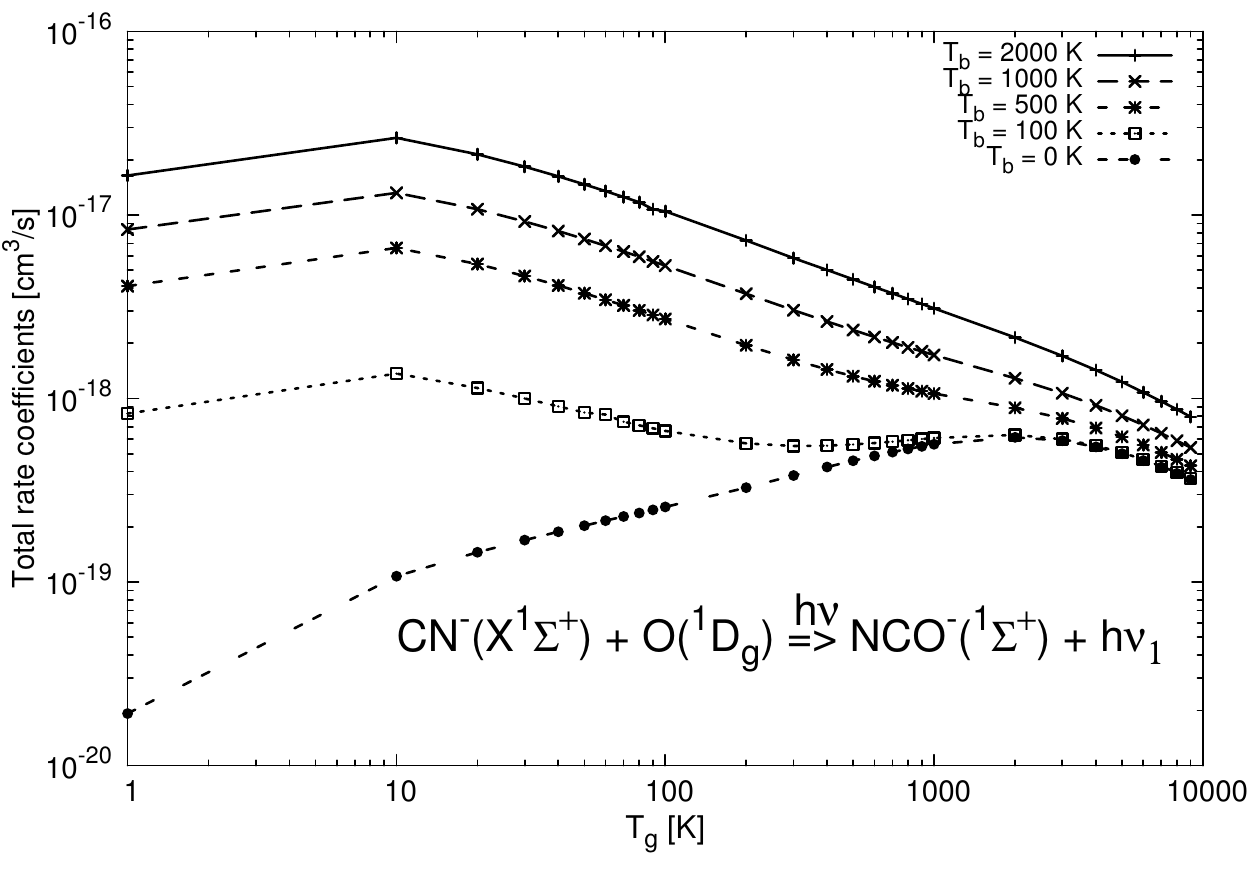}
\caption{Rate coefficients for stimulated and spontaneous radiative association of $\mathrm{NC^-}$ and $\mathrm{O}$ for various blackbody radiation temperatures. The spontaneous process is given by the curve labelled "0 K". }
\end{figure}

\begin{center}
 \begin{tabular}{ | l | l | }
    \hline Ion & Rate [$\mathrm{10^{-18}}$ $\mathrm{cm^3/s}$] \\ \hline
     $\mathrm{C_2H^-}$ & ~~~~$\sim$ 0.01  \\
     $\mathrm{NCO^-}$ & ~~~~$\sim$ 0.25   \\
     $\mathrm{Li^+ +H}$ & ~~~~$\sim$ 0.00018   \\
     $\mathrm{Li +H^+}$ & ~~~~$\sim$ 7660   \\
     $\mathrm{HeH^+}$ & ~~~~$\sim$ 0.02   \\
     $\mathrm{CH^+}$ & ~~~~$\sim$ 50    \\
\hline
    \end{tabular}

\captionof{table}{The comparisons of our computed total rates for the spontaneous RA  with the rates of other molecular cations at a gas temperature of $\mathrm{T_g}$ = 50 K. See main text for the relevant references.}
\end{center}
\clearpage
As already discussed in the present Introduction, another common path to the formation of molecular anions in the ISM environment is the interaction of the corresponding neutral radicals with environmental electrons, a mechanism which can lead to the strongly exothermic process which attaches such electrons to the neutral molecules, thereby producing stable molecular anions [5,33]. In particular the REA mechanisms follow the reactions:

\begin{equation}
C_2H(X ^2\Sigma^+) + e^- \rightarrow C_2H^-(X ^1\Sigma^+) + \hbar\omega
\end{equation}
\begin{equation}
NCO(X ^2\Sigma^+) + e^- \rightarrow NCO^-(X ^1\Sigma^+) + \hbar\omega
\end{equation}

here  the stabilization of either anion occurs by emitting the excess energy radiatively. The crux of the matter with the above process is to accurately evaluate the relative efficiency of the REA stabilization path versus the competing autodetachment channels that will release the excess electrons after temporary attachment into a metastable anion [31,15] : $( M^-)^* \rightarrow M + e^-$. The studies involving the $\mathrm{C_2H^-}$ formation via either direct REA processes as in eq.s (14) and (15), or via indirect REA processes that dissipate the excess energy of its large and positive Electron Affinities (EAs) into the vibrational network of the molecular nuclei [5, 16, 33], indicate that stabilization of the molecular anion, in the case of small molecules,  is an inefficient process with respect to the autodetachment channels: the value of rates produced in ref.[15] for the $\mathrm{C_2H^-}$ system at 30 K is 7x$\mathrm{10^{-17}}$ $\mathrm{cm^3/s}$, which makes the REA process not a very important one for producing the molecular anion as a stable species. In the case of the formation of $\mathrm{NCO^-}$, no direct calculations for the process (15) exist thus far, although the similarities of its physical characteristics (large and positive EA of 3.609 eV, low-density of vibrational modes in a few-atom molecular network) with those of the $\mathrm{C_2H^-}$ also indicate that the corresponding abundances for the stable anion may not be too large. A recent ,detailed search for its presence in several DMC regions [14] turned out to be still inconclusive as to its unequivocal detection. Our present data of Table 3 indeed also suggest for both  systems that the probabilities for the spontaneous formations of either molecular anions are even smaller  than those suggested for the REA mechanisms. Our present result therefore suggests once again that the possible existence of $\mathrm{C_2H^-}$ and $\mathrm{NCO^-}$ is not likely to occur  efficiently via either the RA or the REA dynamic mechanisms of formation.

      It is also interesting to note from our results reported by Figures 13 and 14 that the stimulated formation in energetic photon baths increases the RA rates by several orders of magnitude but never sufficiently to indicate that formation path would become substantially important: the values are similar to those found [16] for the direct REA mechanisms, and those rates were ,deemed to yield fairly low efficiency for that process.

%\clearpage

\section{CONCLUSIONS}

In this work we have analyzed in some detail the formation of $\mathrm{C_2H^-}$ and $\mathrm{NCO^-}$ by computationally exploring the possibility and efficiency of the RA process that takes place  in their ground electronic states. First the potential energy curves for both anions have been obtained using accurate ab initio methods and, as the next step,  we have computed the spontaneous and stimulated radiative association cross sections at various bath temperatures $\mathrm{T_b}$, and for different values of the photon bath in order to treat also the radiation-stimulated process. We have then  calculated the corresponding rate coefficients for the formation of the molecular anions under a wide range of temperatures: from the low values expected in the DMC environments to the higher temperatures of the diffuse regions of the ISM.
 
Although outside the direct aim of the present study, such produced quantitities can be further used to estimate the $\mathrm{C_2H^-}$ and $\mathrm{NCO^-}$ abundances in the ISM and therefore employed within different evolutionary schemes. The latter could then provide the expected level of abundance within different chemical environments.
The radiative path to anionic formation has been explored in this study by first considering each triatiomic system as a pseudo-2Body structure in which the stronger bonds of the anionic partners are kept essentially at their equilibrium value since the DMC environment would have them in their ground rotovibrational states as the most likely states. Furthermore, we have additionally found with our calculations that the collinear approaches by the  neutrals, i.e. by H and O atoms, provide the strongest interaction potentials and the largest dipolar functions with respect to the "bent" approaches. In order to generate upper-limit values of the final rates we have therefore carried out the present study by employing the collinear configurations for the analysis of the RA mechanism.
 
  We have additionally located the presence of several resonant structures ,due to metastable complex formations during collisions, for both systems and showed their effects on the final size of the associative rates.

       We have also analysed the final rates in comparison with the corresponding sizes of similar RA rates insimple ions detected in the ISM: $\mathrm{LiH^+}$ , $\mathrm{HeH^+}$  and $\mathrm{CH^+}$, just to check on a few of the most studied species. The comparison of those rates with our present results indicates that the rates we have found are similar in size to those computed earlier for these ionic systems but are all fairly small with respect to those that would be required to make the process  unequivocally the most important for the formations of the title anions. As a matter of fact, the comparison with the competitive radiative association mechanism involving free electrons, i.e. the REA processes for forming the anions from the neutral radicals, discussed in earlier studies[16,33], indicate that in such molecular species with a small number of atoms, and therefore with a low density of phase-space vibrational states, the dissipation of the large amounts of energy (associated with their large and positive EAs) during  electron attachment is not an efficient mechanism for stabilizing the final bound molecular anions. It  therefore follows that, since  those computed REA rates are comparable in size with the present RA rates, both are  not sufficiently large for prevailing on the detachment paths which cause the loss of the metastable electron: autodetachment and associative detachment mechanisms [16,17].
       It is interesting to further note, in fact, that the laboratory studies on the reactions of $\mathrm{C_2^-}$ with $\mathrm{H}$ atoms [17] at room temperatures have discovered that in such a small C-bearing chain the main reaction product is the neutral species originating from AD mechanism: $\mathrm{C_2H}$  that was produced with a rate of  $\mathrm{7.7x10^{-10}}$ $\mathrm{cm^3/s}$ [17]. This mechanism therefore provides a further indication that yet another possible reaction involving the small polyyne would only lead to the loss of the anionic species discussed in the present study..

      In conclusion, the existing difficulties for the confirmed sighting of either title anion by experimental efforts that we have discussed earlier (e.g. see ref.[14]) are justified here by the  low efficiency found in our calculations for  at least three suggested mechanisms of their formation: the  limited size of  the computed  RA rates,  the comparable smallness of the REA rates suggested in the current literature and the efficency of the anion destruction ( in the case of the $\mathrm{C_2H^-}$ ) by AD reactions with $\mathrm{H}$ atoms.
This therefore suggests that other, more complex chemical routes need to be investigated, one being, for example, the reaction of  neutral acetylene  with $\mathrm{H^-}$ to form $\mathrm{H_2}$ and the corresponding molecular anion, as already discussed by us for the case of the cyanide anion [33], just to quote a possible option. The possible validity of such a suggestion is presently being investigated in our group and will be reported elsewhere in the near future.

  The present work has therefore provided quantitative information from an ab-inito study on the possible efficiency of the RA mechanisms in forming the title anions, thus shedding further light on the ongoing question of the most likely mechanisms of formation of the smaller molecular anions in DMC and in the ISM in general.

\section{Acknowledgments}

The computational support of the Center for Computation of the University of Innsbruck is gratefully acknowledged. One of us (I.I.) is  also indebted with the Marie-Curie Initial Training Network "COMIQ: Cold Molecular Ions at the Quantum limit" for  the awarding of a early Training Fellowship. We also acknowledge the support of the Austrian Science Fund ( FWF), project P27047-N20.

\clearpage

%\begin{center}

\end{document}